\documentclass[11pt]{JHEP3}
\usepackage{amssymb,epsfig,amsmath,amsfonts,graphicx}

\newcommand{\Tr}{\text{\,Tr\,}}
\newcommand{\Str}{\text{\,Str\,}}

\newcommand{\LL}{\mathcal{L}}

\newcommand{\TeV}{\text{ TeV}}
\newcommand{\GeV}{\text{ GeV}}

\newcommand{\half}{\frac{1}{2}}
\newcommand{\hc}{\text{ h.c. }}
\newcommand{\identity}{{\rlap{1} \hskip 1.6pt \hbox{1}}}

\title{A ``Littlest Higgs'' Model with Custodial $SU(2)$ Symmetry}

\author{Spencer Chang\\ 
Jefferson Physical Laboratory, Harvard University, Cambridge, MA 02138\\
email: chang@physics.harvard.edu}

\preprint{hep-ph/0306034\\HUTP-03/A048}

\abstract{
In this note, a ``littlest higgs'' model is presented which has an approximate 
custodial $SU(2)$ symmetry.  The model is based on the coset space $SO(9)/(SO(5)\times SO(4))$.
The light pseudo-goldstone bosons of the theory 
include a {\it single} higgs
doublet below a TeV and a set of three $SU(2)_W$ triplets and an electroweak singlet in the TeV range.
All of these scalars obtain approximately 
custodial $SU(2)$ preserving vacuum expectation values.  This model addresses a defect
in the earlier $SO(5)\times SU(2)\times U(1)$ moose model, with the only extra complication being an
extended top sector.  Some of the precision electroweak observables are 
computed and do not deviate appreciably from Standard Model predictions.  In an S-T oblique analysis,
the dominant non-Standard Model contributions are the extended top sector and higgs doublet contributions.
In conclusion, a wide range of higgs masses is allowed in a large region of parameter space consistent with 
naturalness, where large higgs masses requires some mild custodial $SU(2)$ violation from the 
extended top sector.
}  


\begin{document}
\section{Introduction}
In the near future, experimental tests at the LHC 
will begin to map out physics at the TeV energy scale.  With this data, a determination of the
higgs sector, and more importantly, discovering the physics that stabilizes the weak scale from radiative
corrections should be achievable goals.  However, in the interim,
the industry of precision electroweak observables has given us some indirect evidence on
what the theory beyond the standard model must look like. 
And given the unreasonably
good fit of the standard model to these observables, these constraints generically suggest
 a theory with perturbative physics at the TeV scale.  

For many years, the only models that could stabilize the weak scale and be weakly perturbative
were supersymmetric models, most notably the MSSM.  In the past two years, it has been 
shown that there is a new class of perturbative theories of electroweak symmetry breaking, 
that of the ``little higgs'' \cite{Arkani-Hamed:2001nc, Arkani-Hamed:2002pa,Arkani-Hamed:2002qx,
Gregoire:2002ra, Arkani-Hamed:2002qy, Low:2002ws, Kaplan:2003uc, Chang:2003un, Skiba:2003yf}.  For reviews of the physics, 
see \cite{Wacker:2002ar, Schmaltz:2002wx} and for more detailed phenomenology see
\cite{Burdman:2002ns, Han:2003wu, Dib:2003zj, Han:2003gf}.  Little Higgs theories protect the higgs boson
from one-loop quadratic divergences because each coupling treats the higgs
boson as an exact goldstone boson.  However,  two different couplings together 
can break the non-linear symmetries protecting the higgs mass, and thus
the higgs is a pseudo-goldstone boson with quadratic divergences to its mass 
pushed to two-loop order.  This allows a separation of scales between the cutoff and the
electroweak scale, so that physics can be perturbative until the cutoff is reached at $\Lambda
\approx 10$ TeV.

Having weakly perturbative physics at the TeV scale is probably necessary but definitely not
sufficient to guarantee a theory is safe from precision electroweak constraints.  Currently precision
observables have been measured beyond one-loop order in the standard model, and since 
little higgs model corrections are parameterically of this order, these observables can put 
constraints on these theories \cite{Chivukula:2002ww, Hewett:2002px, Csaki:2002qg,Csaki:2003si,Kribs:2003yu,Gregoire:2003kr}.
However, these constraints are not unavoidable, and isolating the strongest constraints 
can point to the necessary features to make little higgs models viable theories
of electroweak symmetry breaking.  First of all, there are modifications of the original models
which address these strongest constraints and 
greatly ammeliorate the issue \cite{Csaki:2003si}.  However, just recently, a little higgs model was introduced 
containing a custodial $SU(2)$ symmetry, the $SO(5)\times SU(2)\times U(1)$ moose model
\cite{Chang:2003un}.  In the limit of strong coupling for the $SO(5)$ gauge group, the 
precision electroweak constraints due to the T parameter were softened and in general, there
is a large region of parameter space consistent with precision electroweak constraints
and naturalness \cite{Csaba:talk}.

Let's briefly summarize the physics that gives the custodial $SU(2)$ symmetry.  The important
point is that in models with a gauged $U(1)\times U(1)$ subgroup and standard model fermions
gauged under just one of the $U(1)$'s, 
the massive $B'$ of these theories provides two constraints.  The first constraint is that
integrating out the
$B'$ generates a custodial $SU(2)$ violating operator that 
after electroweak symmetry breaking corrects the standard model formula 
for the mass of the $Z$ gauge boson.  This gives corrections
to the $\rho$ parameter, and vanishes as the two $U(1)$ gauge couplings become equal.   
However, the second constraint pulls in the opposite direction in gauge parameter space.
This is because the coupling of the $B'$ to standard model fermions generates corrections
to low energy four-fermi operators and also to coefficients of the $SU(2)_W\times U(1)_Y$ 
fermion currents.  These corrections vanish in the limit in which the $U(1)$ that the standard model
fermions is not gauged under becomes strong.  Thus, these two constraints prefer different
limits in parameter space and can constrain the model.

As pointed out before \cite{Chang:2003un}, there are simple modifications that evade these 
two constraints, such as only gauging $U(1)_Y$, charging the SM fermions equally under both
$U(1)$'s, or through fermion mixing.  Another simple approach that gives custodial $SU(2)$ symmetry
is to complete the $B'$ into a custodial $SU(2)$ triplet.  If the triplet is exactly degenerate in mass, 
integrating it out
does not contribute to a custodial $SU(2)$ violating operator.
To include these new states, instead of gauging two $U(1)$'s, 
$SU(2)_R\times U(1)$ is gauged. After being broken down to the diagonal $U(1)_Y$,  
$B'$ and $W^{r\,\pm}$ are put into a ``$SU(2)_R$'' triplet.  Integrating out the $W^{r\,\pm}$ generates
an operator which only gives mass to the $W$ giving a $\rho$ contribution of the opposite sign
of the $B'$ contribution.  Numerically, the total $\rho$ contribution from the 
gauge sector cancels in the strong $SU(2)_R$ coupling limit (where the triplet becomes
degenerate), which is the same limit
that reduces corrections to fermion operators.

However, this cancelation is not quite exact for the $SO(5)\times SU(2)\times U(1)$ moose
model.
The higgs quartic potential of that theory has a flat direction when the two higgs vevs have
the same phase, thus viable electroweak symmetry breaking requires the higgs vevs to have
different phases.  
This phase difference changes the $\rho$ contribution due to the $W^{r\,\pm}$ gauge bosons.
The higgs currents of the $W^{r\,\pm}$ are not invariant under a vev phase
rotation, and thus the cancelation in the
strong coupling limit only occurs if the phase is 0 or $\pi$.  Indeed, this remnant of custodial $SU(2)$ 
violation puts the strongest constraint on the theory.

The situation can be easily resolved if the little higgs theory  contains only a single light higgs doublet.
In this case, the $W^{r\,\pm}$ current just transforms
by a phase under the vev phase rotation, which cancels out of the contribution.  
It turns out that the $SO(5)\times SU(2)\times U(1)$ moose's defect can be removed by 
imposing a $\mathbb{Z}_4$ symmetry inspired by orbifold models \cite{Wacker:toappear}, 
which leaves only a single light
higgs doublet that still has an order one quartic coupling.  In this paper, we will take a different
approach and construct a ``littlest higgs'' model with custodial $SU(2)$ symmetry and 
just {\em one} higgs doublet.    

This ``littlest higgs'' model will be based on an $\frac{SO(9)}{SO(5)\times SO(4)}$ coset
space, with an $SU(2)_L\times SU(2)_R \times SU(2) \times U(1)$ subgroup of $SO(9)$ gauged.
The pseudo-goldstone bosons are a single higgs doublet, an electroweak singlet and a set of
three $SU(2)_W$ triplets, precisely the content of one of the original custodial $SU(2)$
preserving composite higgs models \cite{Georgi:1984af}.
The global symmetries 
protect the higgs doublet from one-loop quadratic divergent contributions to its mass.  However,
the singlet and triplets are not protected, and will be pushed to the TeV scale.  
Integrating
out these heavy particles will generate an order one quartic coupling for the higgs.
To complete the theory with fermions, 
the minimal top sector contains two extra colored quark doublets and their charge
conjugates.

Since the primary motivation of the model is to improve consistency with precision electroweak
observables, the model's corrections to these observables will be calculated.  First, we will
see that aside from some third generation quark effects, a limit will exist where non-oblique 
corrections vanish.  This limit was recently described as ``near-oblique'' \cite{Gregoire:2003kr}
and we will continue to use this terminology.  The existence of this limit allows a meaningful
S and T analysis of the oblique corrections, which will be performed in this model to 
order $(v^2/f^2)$.  The dominant contributions come from the extended top sector and the 
higgs doublet, which are quite mild in most of parameter space.  
In fact, this analysis will show that there is a wide range of higgs masses allowed
in a large region of parameter space consistent with naturalness.   

The outline of the rest of the paper is as follows:  in section \ref{Sec: Model} we describe 
the model's coset space, light scalars and the symmetries that protect the higgs mass. 
We also analyze the gauge structure   
and then describe the minimal candidate top sectors. In section \ref{Sec: Potential}, 
we will show how the quartic higgs potential is generated as well as describe the log
enhanced contributions to the higgs mass parameter.  There will be 
vacuum stability issues, and we will point out ways which these can be resolved.  
Also as usual, the top sector contributions will generically
drive electroweak symmetry breaking.  In section \ref{Sec: PEWO}, some precision electroweak
observables will be calculated and the constraints on the theory will be detailed.  
In section \ref{Sec: Conclusion}, we conclude and finally in appendix \ref{Sec: Gens}, we
describe our specific generators and representations of $SO(4)$. 
\section{The Model}
\label{Sec: Model}
The first ingredient necessary for custodial $SU(2)$ symmetry is the breakdown of 
$SU(2)_L \times SU(2)_R \times SU(2) \times U(1)$ down to the diagonal 
$SU(2)_W \times U(1)_Y$ subgroup.  Therefore the global symmetry group must be at
least rank 4.  Two rank 4 groups are easy to eliminate-- $SU(5)$ does not contain the gauged group and the $SO(8)$
adjoint contains no higgs doublets.  
This leaves $SO(9)$, $Sp(8)$ and $F_4$  as the only remaining rank 4
candidates. In this paper, we'll focus on the  $SO(9)$ group as it is the easiest to analyze.
However, we do mention here that it appears to be difficult to get a single light higgs doublet 
in the $Sp(8),F_4$ groups.

Isolating our attention to $SO(9)$, it is straightforward to implement the ``little higgs''
construction.  Using the vector representation,
the top four by four block will contain the gauged $SO(4) \cong SU(2)_L \times SU(2)_R$ 
and the bottom four by four block will contain the gauged $SU(2)\times U(1) \subset SO(4)$.
The coset space should break these two $SO(4)$'s down to their diagonal subgroup, which 
can be achieved by an off-diagonal vev for a two-index tensor of $SO(9)$.  In order to 
have the largest unbroken global symmetry (and thus reduce the amount of light scalars),
a symmetric two-index tensor should be chosen.

This construction can be described in the following way:  take an orthogonal 
symmetric nine by nine matrix, representing a non-linear sigma model field $\Sigma$ 
which transforms under an $SO(9)$ rotation
by $\Sigma \to V\Sigma V^T$.  To break the $SO(4)$'s to their diagonal, we take $\Sigma$'s
vev to be 
\begin{eqnarray}
\langle \Sigma \rangle=\left(\begin{array}{ccc}
0 & 0 & \identity_4 \\
0 & 1 & 0 \\
\identity_4 & 0 & 0 \end{array}\right)
\end{eqnarray}
which breaks the $SO(9)$ global symmetry down to an $SO(5)\times SO(4)$ subgroup.\footnote{
We could separate the trace from $\Sigma$ to make it transform as an irreducible representation
of $SO(9)$, however this equivalent vev is chosen so that $\Sigma$ can be orthogonal.
}
This coset space guarantees the existence of $20 = (36-10-6)$ light scalars. Of these
20 scalars, 6 will be eaten in the higgsing of the gauge groups down to $SU(2)_W \times U(1)_Y$.
The remaining 14 scalars consist of a single higgs doublet $h$, an electroweak singlet $\phi^0$,
and three triplets $\phi^{ab}$ which transform under the $SU(2)_L \times SU(2)_R$ diagonal
symmetry as\footnote{See appendix \ref{Sec: Gens} for specific representation and generator conventions.}
\begin{eqnarray}
h: (\mathbf{2}_L,\mathbf{2}_R) \hspace{0.5in} \phi^0 : (\mathbf{1}_L,\mathbf{1}_R)
\hspace{0.5in} \phi^{ab} : (\mathbf{3}_L,\mathbf{3}_R).
\end{eqnarray}
This spectrum is particularly nice as each set of scalars can have vacuum expectation values
that preserve custodial $SU(2)$; we will see later that this is approximately true.
These fields parameterize the direction of the $\Sigma$ field and can be written in the 
standard way
\begin{eqnarray}
\Sigma = e^{i\Pi/f}\langle\Sigma\rangle e^{i\Pi^T/f} = e^{2i\Pi/f}\langle\Sigma\rangle 
\end{eqnarray}
where 
\begin{eqnarray}
\Pi = \frac{-i}{4}\left(\begin{array}{ccc} 
	0 & \sqrt{2} \vec{h} & -\Phi \\
	-\sqrt{2} \vec{h}^T & 0 & \sqrt{2} \vec{h}^T \\
	\Phi & -\sqrt{2} \vec{h} & 0
\end{array}\right). 
\end{eqnarray}
In $\Pi$, the would-be goldstone bosons that are eaten in the higgsing down 
to $SU(2)_W\times U(1)_Y$ have been set to zero.  The singlet and triplets are
contained in the symmetric four by four matrix $\Phi$ where
\begin{eqnarray}
\Phi = \phi^0 + 4\phi^{ab}\: T^{l\,a} T^{r\,b}.
\end{eqnarray}

It is now simple to determine the global symmetries that protect the higgs
mass at one loop.  Under the upper five by five $SO(5)_1$
symmetry, the scalars transform as:
\begin{eqnarray}
\delta \vec{h} = \vec{\alpha} + \cdots \hspace{0.4in} 
\delta \Phi = -\frac{1}{2f}\left(\vec{\alpha}\: \vec{h}\,^T + \vec{h}\: \vec{\alpha}\,^T\right) +\cdots
\end{eqnarray}
Similarly, under the lower five by five $SO(5)_2$, the scalars transform as:
\begin{eqnarray}
\delta \vec{h} = \vec{\beta} + \cdots \hspace{0.4in} 
\delta \Phi = \frac{1}{2f}\left(\vec{\beta}\: \vec{h}\,^T + \vec{h}\: \vec{\beta}\,^T\right) + \cdots
\end{eqnarray}
Any interaction that preserves at least one of these $SO(5)$ symmetries treats the higgs as
an {\em exact} goldstone boson.  Thus, if all interactions are chosen to preserve one of these
symmetries, the higgs mass will be protected from one loop quadratic divergences.  In the
next two subsections, that motivation is used to determine the requisite interactions of 
the theory.  
\subsection{Gauge Sector}
\label{Sec: Gauge}
The gauge group structure obviously follows the preserving $SO(5)$ symmetry logic.
The gauged $SO(4) \cong SU(2)_L\times SU(2)_R$ is generated by
\begin{eqnarray}
\tau^{l\,a} = \left(\begin{array}{cc} T^{l\,a}& \\ &0_5 \end{array}\right) \hspace{0.3in}
\tau^{r\,a} = \left(\begin{array}{cc} T^{r\,a}& \\ &0_5 \end{array}\right)
\end{eqnarray}
and preserves $SO(5)_2$ whereas the gauged $SU(2)\times U(1)$ is generated by
\begin{eqnarray}
\eta^{l\,a} = \left(\begin{array}{cc} 0_5& \\ &T^{l\,a}  \end{array}\right) \hspace{0.3in}
\eta^{r\,3} = \left(\begin{array}{cc} 0_5& \\ &T^{r\,3} \end{array}\right)
\end{eqnarray}
and preserves $SO(5)_1$.
The kinetic term for the pseudo-goldstone bosons can now be written as 
\begin{eqnarray}
\LL_{kin} = \frac{f^2}{4}\Tr\left[D_\mu\Sigma D^\mu\Sigma\right] 
\end{eqnarray}
where the covariant derivative is given by
\begin{eqnarray}
D_\mu \Sigma = \partial_\mu\Sigma +i\left[A_\mu,\Sigma \right] 
\end{eqnarray}
with the gauge boson matrix $A_\mu$ defined by 
\begin{eqnarray}
A \equiv g_L W^{la}_{SO(4)} \tau^{l\,a} + g_R W^{ra}_{SO(4)} \tau^{r\,a} + g_2  W^{la} \eta^{l\,a}+ g_1 W^{r3} \eta^{r\,3}. 
\end{eqnarray}

Due to the vev of $\Sigma$,  
the vector bosons mix and can be diagonalized with the following
transformations:
\begin{eqnarray}
\nonumber
B &=& \cos\theta' W^{r3} - \sin\theta' W_{SO(4)}^{r3}
\hspace{0.5in}
B'= W'{}^{\,r3} = \sin \theta' W^{r3} + \cos \theta'  W_{SO(4)}^{r3}\\
W^a &=& \cos\theta W^{la} - \sin\theta W_{SO(4)}^{la}
\hspace{0.5in}
W'{}^a=W'{}^{\,la} = \sin \theta W^{la} + \cos \theta W_{SO(4)}^{la}
\end{eqnarray}
where the mixing angles are related to the couplings by:
\begin{eqnarray}
\nonumber
\cos \theta' &=& g'/g_1 \hspace{0.5in} \sin \theta' = g'/g_R \\
\cos \theta &=& g/g_2 \hspace{0.5in} \sin \theta = g/g_L. 
\end{eqnarray}
Notice that there is no relation between $\theta$ and $\theta'$ since 
$SO(4)$ has two arbitrary gauge couplings $g_L$ and $g_R$.  They could of course 
be set equal by imposing a $\mathbb{Z}_2$ symmetry, which we will choose to do when describing
the limits on the model.  In this L-R symmetric limit, the constraint on the angles in
order to get the correct $\theta_W$ is
$\sin \theta \approx \sqrt{3} \sin \theta'$. 
The masses for the heavy vectors can now be written in terms of the electroweak gauge
couplings and mixing angles:
\begin{eqnarray}
m^2{}_{W'}  =  \frac{4g^2 f^2}{\sin^2 2\theta} 
\hspace{0.5in}
m^2{}_{B'} =  \frac{4g'{}^2 f^2}{\sin^2 2 \theta'} 
\hspace{0.5in}
m^2{}_{W^{r\,\pm}} = \frac{4g'^2 f^2}{\sin^2 2 \theta'} \cos^2 \theta'.
\end{eqnarray}

\subsection{Fermion Sector}
\label{Sec: Fermion}
For all fermions besides the top quark, the yukawa couplings are small, and thus it is not
necessary to protect the higgs from their one loop quadratic divergences.  However, there is
the requirement that low energy observables such as four-fermi operators do not receive
large corrections.  This  can be achieved by gauging the light fermions only under $SU(2)\times U(1)$.
In the strong $SO(4)$ coupling limit, these fermions will decouple from the $W'$ and $B'$ 
and will not give strong precision electroweak corrections.

To implement the Yukawa couplings for the light fermions, we add
\begin{eqnarray}
\LL_{LF} = \sqrt{2}f\left[y_u\left(0_4 \, u^c \, 0_4\right)\Sigma\left(\begin{array}{c}
0_5 \\ \vec{q}_u\end{array}\right) +
y_d\left(0_4 \, d^c \, 0_4\right)\Sigma\left(\begin{array}{c}
0_5 \\ \vec{q}_d\end{array}\right) +
y_l\left(0_4 \, e^c \, 0_4\right)\Sigma\left(\begin{array}{c}
0_5 \\ \vec{l}\end{array}\right)
\right] +\hc
\nonumber
\\
\end{eqnarray}
In this expression, we have defined the ``$SO(4)$'' representations corresponding to 
the $SU(2)\times U(1)$ representations by 
\begin{eqnarray}
\vec{q}_u \leftrightarrow Q_u = \left(q \; 0_2\right) \hspace{0.45in}
\vec{q}_d \leftrightarrow Q_d = \left(0_2 \; q\right) \hspace{0.45in}
\vec{l} \leftrightarrow L = \left(0_2 \; l\right) 
\end{eqnarray}
where $q$ and $l$ are the standard quark and lepton doublets.  The exact correspondence
between the two equivalent representations is presented in appendix \ref{Sec: Gens}.  
At first order, these interactions reproduce the standard yukawa interactions for the
light fermions.

On the other hand, the top yukawa is the strongest one loop quadratic divergence of the
standard model and therefore the top sector must be extended in order to stabilize the 
higgs mass parameter.  From the symmetry considerations given earlier, the top sector has 
to preserve either the $SO(5)_1$ or $SO(5)_2$ symmetry.  The minimal approach is to 
preserve the $SO(5)_1$ symmetry, which can be accomplished by adding $t^c$
to an $SO(4)$ gauge vector $\mathcal{\vec{X}}^c$.  In addition to this new vector, we add its 
charge conjugate $\mathcal{\vec{X}}$ and add a Dirac mass for the two fermions.   
The interactions are:
\begin{eqnarray}
\LL_{top} = y_1 f \:(\vec{\mathcal{X}}^c\,^T \; t^c \; 0_4)\,\Sigma
   \left(\begin{array}{c} 0_5 \\ \vec{q}_t\end{array}\right)
+y_2 f \vec{\mathcal{X}}\,^T \vec{\mathcal{X}}^c + \hc
\end{eqnarray}
Now, the choice is whether or not to make $\vec{q}_t$ a ``full'' $SO(4)$ vector.  Since it is only
charged under $SU(2)\times U(1)$, it does not have to be a full $SO(4)$ vector, but can 
contain just one doublet like $\vec{q}_u$ above.  For the sake of simplicity, we will choose to 
analyze the most minimal case of one doublet. 

In this minimal case, the gauge charges of the fermions are:
\begin{eqnarray}
\begin{array}{c|ccccr}
 & SU(3)_c & SU(2)_L & SU(2)_R & SU(2) & U(1) \\
\hline
q   & 3 &1 & 1 & 2 & 1/6 \\
t^c & \bar{3} &1 & 1 & 1 & -2/3 \\
\vec{\mathcal{X}} & 3 & 2 & 2 & 1 & 2/3 \\
\vec{\mathcal{X}}^c & \bar{3} &2 & 2 & 1 & -2/3
\end{array}
\end{eqnarray}
Under the diagonal $SU(2)_W\times U(1)_Y$, $\vec{\mathcal{X}}$ contains two doublets $\mathcal{X}_1,
\mathcal{X}_2$ with hypercharge $1/6$ and $7/6$ respectively.  Expanding the terms, we find
a mass term linking $\mathcal{X}_1^c$ with a linear combination of $q$ and $\mathcal{X}_1$.
Integrating out the heavy fermion gives a top yukawa coupling 
\begin{eqnarray}
y_t = \frac{y_1y_2^*}{\sqrt{2(|y_1|^2+|y_2|^2)}}.
\end{eqnarray}

\section{Potential and EWSB breaking}
\label{Sec: Potential}
By construction, the interactions of the theory do not generate one loop quadratic divergences
for the mass parameter of the higgs.  To demonstrate this explicitly,  the 
Coleman-Weinberg potential will be computed.  The one loop quadratic divergent piece 
will generate a potential for $\Phi$ and $h$, including a quadratically divergent mass
for the $\Phi$.  Similar to the $SU(6)/Sp(6)$ model \cite{Low:2002ws}, the gauge interactions 
will introduce an instability in the vacuum.  The problem is a bit more serious here
because the gauge contributions are opposite in sign for the singlet and triplet masses;
thus, the origin of the potential is a saddle point.  However, as in the $SU(6)/Sp(6)$ paper,
there are ways to cure this instability issue.  Once the instability has been addressed, 
integrating out the massive $\Phi$ will generate an order one quartic coupling for 
$h$, but no mass term.  

For the log divergent piece of the Coleman-Weinberg potential, 
we will only analyze the contributions to the higgs mass parameter.  As usual, gauge and
scalar sectors will give positive contributions whereas the top sector gives a large negative
contribution that drives electroweak symmetry breaking.
\subsubsection*{One Loop Quadratic Term}
The one loop quadratically divergent piece of the Coleman-Weinberg Potential is given by
\begin{eqnarray}
V_\text{one loop $\Lambda^2$}=\frac{\Lambda^2}{32\pi^2}\Str(M^\dag M[\Sigma])
\end{eqnarray}
By the symmetry arguments given earlier, the different $SO(5)_i$ preserving 
interactions can generate operators depending on   
\begin{eqnarray}
SO(5)_1  :\quad \Phi +\frac{1}{2f} \vec{h}\: \vec{h}\,^T
\hspace{0.45in} \text{or} \hspace{0.45in}
SO(5)_2  :\quad \Phi -\frac{1}{2f} \vec{h}\: \vec{h}\,^T.
\end{eqnarray}
It will also be convenient to introduce some notation, where 
\begin{eqnarray}
\frac{1}{2f} \vec{h}\: \vec{h}\,^T = \mathcal{H}^0 + 4\mathcal{H}^{ab}\: T^{l\,a} T^{r\,b}.
\end{eqnarray}
$\mathcal{H}^0$ and $\mathcal{H}^{ab}$ are quadratic in the $h$ fields and their explicit
expressions appear in appendix \ref{Sec: Gens}. 

The gauge contribution can be calculated from the kinetic term for $\Sigma$, which gives
\begin{eqnarray}
\nonumber
V_\text{gauge} =  && -\frac{9f^2}{8}\left[(g_L^2+g_R^2)
(\phi^0-\mathcal{H}^0)^2  +(g_2^2+g_1^2/3) 
(\phi^0+\mathcal{H}^0)^2\right]
+ \\  & &
\frac{3f^2}{8}\left[(g_L^2+g_R^2) (\phi^{ab}-\mathcal{H}^{ab})^2
 + (g_2^2+g_1^2)
(\phi^{ab}+\mathcal{H}^{ab})^2-2g_1^2(\phi^{a3}+\mathcal{H}^{a3})^2
\right]
\end{eqnarray}
where we have ignored a constant term, expanded to 
second order in $\Phi$ and fourth order in $h$, and set $\Lambda = 4\pi f$.  
There are 
two important points to make about this result.  First of all, there is a sign difference
between the mass terms for the singlet and triplets.  Thus, the gauge interactions
introduce a saddle point instability in the vacuum.  
This is expected since the gauge groups would prefer the  $\Sigma$ vev to
be proportional to the identity; at this vacuum, no gauge groups are broken and indeed
the negative mass squared for the singlet attempts to rotate the vev to this non-breaking
vacuum.  However, as we will see later, the top sector gives equal sign contributions to both
mass terms. Also from the point of view of the effective field theory, operators can be
written down that give equal sign contributions to both masses or even just to the singlet. 
The second thing to note about the gauge contribution is that only the gauged $U(1)$ 
introduces explicit custodial $SU(2)$ violation into the potential.  As a matter of fact,
this will be the only interaction that can give the triplets a custodial $SU(2)$ violating
vev.  Since $g_1$ will be approximately equal to the standard model 
hypercharge coupling, the triplet vevs usually give suitably small 
contributions to $\rho$.  We will analyze the triplet vevs in greater detail in section 
\ref{Sec: PEWO}.   

Now, analyzing the top sector, we find the contribution 
\begin{eqnarray}
V_\text{fermion} =   3|y_1|^2f^2\left[
(\phi^0+\mathcal{H}^0)^2  
 + 
(\phi^{ab}+\mathcal{H}^{ab})^2 \right]
\end{eqnarray}
where again we have ignored a constant piece and set $\Lambda = 4\pi f$.  As noted earlier,
the fermion sector gives equal sign contributions to singlet and triplet masses and does
not introduce custodial $SU(2)$ breaking at this order.\footnote{If we had chosen
$\vec{q}_t$ to contain 
two doublets, there would be no custodial $SU(2)$ breaking at any order.}  Following
the $SU(6)/Sp(6)$ little higgs \cite{Low:2002ws}, we could also extend the top sector with
an interaction that preserves the $SO(5)_2$ symmetry.  This would have the added benefit of
giving equal sign contributions to
the $(\phi-\mathcal{H})^2$ terms and could lift the saddle point into a local 
minimum.  Another way
to do this is through operators such as 
\begin{eqnarray}
\LL_1 = a_1 f^2 \sum_{i=1,j=1}^{4,4}\Sigma_{ij}\Sigma_{ij} = a_1 f^2\left[
(\phi^0-\mathcal{H}^0)^2  
 + 
(\phi^{ab}-\mathcal{H}^{ab})^2 \right]
\end{eqnarray}
or 
\begin{eqnarray}
\LL_2 = a_2 f^2 (\sum_{i=1}^{4}\Sigma_{ii})^2 = 4a_2 f^2\left[
(\phi^0-\mathcal{H}^0)^2\right]
\end{eqnarray}
which respect the $SO(5)_2$ symmetry and give contributions to both singlet and triplet masses
or just masses for the singlets.  Depending on the UV completion of the model, these operators
can be generated; for instance, they might appear naturally in an extended technicolor like 
completion.   

These radiative corrections tell us that we must put in these operators with coefficients
of their natural size of the form  
\begin{eqnarray}
\nonumber
V =&& \lambda_{\mathbf{1}}^- f^2 (\phi^0-\mathcal{H}^0)^2  + \lambda_\mathbf{1}^+ f^2 (\phi^0+\mathcal{H}^0)^2+
  \\ && \lambda_\mathbf{3}^- f^2 (\phi^{ab}-\mathcal{H}^{ab})^2 + \lambda_\mathbf{3}^+ f^2(\phi^{ab}+\mathcal{H}^{ab})^2
+\Delta \lambda_\mathbf{3} f^2 (\phi^{a3}+\mathcal{H}^{a3})^2.
\label{eq: Potential}
\end{eqnarray}
As mentioned before, since $g_1$ will be small, $\Delta \lambda_\mathbf{3} \ll \lambda_\mathbf{3}^\pm$ 
is expected, which 
leads to approximately custodial $SU(2)$ preserving triplet vevs.
We will assume that the singlet and triplet masses are positive; integrating out these 
heavy particles then leads to a quartic coupling of the higgs (ignoring $\Delta \lambda_\mathbf{3}$ for 
simplicity):
\begin{eqnarray}
\lambda |h|^4 \quad \text{where} \quad 4\lambda = \lambda_\mathbf{1} + 3\lambda_\mathbf{3}
\end{eqnarray}
and we've defined $1/\lambda_\mathbf{(1,3)} = 1/\lambda_\mathbf{(1,3)}^- +1/\lambda_\mathbf{(1,3)}^+$.
Requiring a positive order one $\lambda$ puts some mild constraints on the $\lambda_{(\mathbf{1},\mathbf{3})}^\pm$ 
parameters.
\subsubsection*{Log Contributions to the Mass Parameter}
Even though the little higgs mechanism protects the higgs from one-loop quadratic divergences,
there are finite, one loop logarithmically divergent, and two loop quadratically 
divergent mass contributions, all of the same order of magnitude.  Here we will analyze
the logarithmically enhanced pieces as given by the
one loop log term in the Coleman-Weinberg 
potential
\begin{eqnarray}
V_\text{one loop log} = \frac{1}{64\pi^2}\Str\left[(M^\dag M)^2 \ln\frac{M^\dag M}{\Lambda^2}
\right].
\end{eqnarray}
The gauge contribution to the mass squared is positive
\begin{eqnarray}
m^2_\text{gauge} = \frac{3}{64\pi^2}\left[3g^2 m^2_{W'} \ln \frac{\Lambda^2}{m^2_{W'}}
+g'^2 m^2_{B'} \ln \frac{\Lambda^2}{m^2_{B'}}\right] 
\end{eqnarray}
but the fermion contribution is negative
\begin{eqnarray}
m^2_\text{fermion} = -\frac{3|y_t|^2}{8\pi^2} m^2_{t'} \ln \frac{\Lambda^2}{m^2_{t'}} 
\end{eqnarray}
where we have defined 
$m^2_{t'} = (|y_1|^2+|y_2|^2)f^2$.\footnote{The heavy $t'$ quark is the only heavy
quark whose mass shifts when the higgs vev is turned on.  Thus, it is the new heavy state that
appears and cuts off the top yukawa quadratic divergence, which is also why the fermionic
contribution to the higgs mass parameter only depends on $m_{t'}$.}  This large top contribution 
generically dominates and drives electroweak symmetry breaking.
We've chosen not to consider the scalar contribution  
since it depends on the
specifics behind the generation of the potential (Eq. \ref{eq: Potential}).
However, we mention that it is typically positive and subdominant to the fermion 
contribution.
 
\section{Precision Electroweak Observables}
\label{Sec: PEWO}
Now that we have described the model's content and interactions, the contributions
to electroweak observables can be calculated.  
In general, we will work to leading order in $O(v^2/f^2)$ and neglect any higher order
effects.
First in section \ref{sec: currents}, 
we will focus on non-oblique corrections to electroweak fermion currents and four-fermi
interactions.  We will demonstrate how the limit of strong $g_L,g_R$ 
coupling is a ``near-oblique'' limit as discussed recently in \cite{Gregoire:2003kr}. 
This limit validates the usefulness of an S and T analysis and in sections 
\ref{sec: tparameter} and \ref{sec: sparameter} we will calculate the model's contributions
to these parameters.  We will choose to keep the S and T contributions from the Higgs sector,
but will subtract out all other standard model contributions.
Finally in section \ref{Sec: Summary}, the results of the full S-T analysis will be presented.

\subsection{Electroweak Currents}
\label{sec: currents}
First of all, there are non-oblique 
corrections due to the exchange of the heavy gauge bosons.  
Specifically, integrating out
the heavy vectors generates four Fermi operators and Higgs-Fermi current current interactions 
(the Higgs-Higgs interactions give oblique corrections and will be
considered in the higgs contribution to T in section \ref{Sec: Vector Bosons}).
The former are constrained by tests of compositeness and the latter 
after electroweak symmetry breaking induce corrections to standard model
fermionic currents which are constrained by Z-pole observables.  
As pointed out recently by Gregoire, Smith, and Wacker \cite{Gregoire:2003kr}, 
the S and T analysis is reliable  
when there exists a ``near-oblique'' limit where most of the  
non-oblique corrections vanish.  The limit
is called ``near-oblique'' since third generation quark physics still has non-vanishing 
effects.  
In this model this limit turns out to be the strong $g_L,g_R\to \infty$ limit
that decouples the light generations from the heavy gauge bosons. 
A discussion of the non-decoupling third generation effects is outside the scope of 
this paper and thus they will not be analyzed. However, for 
a preliminary discussion of the important 
operators in such an analysis, see reference  \cite{Gregoire:2003kr}. 

To calculate these induced effects, we first write down the relevant 
currents to the heavy gauge bosons starting with the higgs (leaving off Lorentz indices
for readability)
\begin{eqnarray}
\nonumber
j^{a}_{W'H} & = &  g \cot 2\theta j_{H}^{a}
= \frac{g \cos{2\theta}}{2 \sin 2\theta} 
\;i h^\dag \sigma^a \overleftrightarrow{D} h\\
j_{B'H} & = &g' \cot 2 \theta' j_{H}
= -\frac{g' \cos{2\theta'}}{2 \sin 2\theta'} 
\;i h^\dagger \overleftrightarrow{D} h
\end{eqnarray}
and also for the standard model fermions (aside from the third generation quarks)
\begin{eqnarray}
j^{a}_{W'F}=g \tan \theta \; j_{F}^a
\hspace{0.5in} j_{B'F}= g'\tan \theta'\; j_{F}
\end{eqnarray}
where they are given in terms of the standard model $SU(2)_W, U(1)_Y$ currents
$j_{(HF)}^a$ and $j_{(HF)}$.  The one heavy gauge boson current left out is the higgs
current to $W^{r\,\pm}$, but since there is no corresponding fermionic current, integrating
out $W^{r\,\pm}$ does 
not generate four-fermi operators or standard model current corrections.

Integrating out the heavy gauge bosons generates the Higgs-Fermi interactions
\begin{eqnarray}
\nonumber
\LL_{\text{H F}} &=& 
-\frac{j_\mu^a{}_{W' \text{H}}\; j^\mu{}^a{}_{W' \text{F}}}{M^2_{W'}}
-
\frac{j_\mu{}_{B' \text{H}}\; j^\mu{}_{B'\text{F}}}{M^2_{B'}}\\
&=&-   \frac{\sin^2 \theta \cos 2\theta}{2 f^2} 
j_{\text{H}}{}^{a\,\mu} j_{\text{F}}{}_{a\,\mu} 
-   \frac{\sin^2 \theta' \cos 2\theta'}{2 f^2} 
j_{\text{H}}{}^{\mu} j_{\text{F}}{}_{\mu} 
\end{eqnarray}
and the four Fermi interactions
\begin{eqnarray}
\nonumber
\LL_{\text{F F}} &=&
- \frac{(j_\mu^a{}_{W'\text{F}})^2}{2 M^2_{W'}}
- \frac{(j_\mu{}_{B'\text{F}})^2}{ 2 M^2_{B'}}\\
&=&
- \frac{\sin^4\theta}{2 f^2}
j_{\text{F}}{}^{a\,\mu} j_{\text{F}}{}_{a\,\mu}
-\frac{\sin^4\theta'}{2 f^2}
j_{\text{F}}{}^{\mu} j_{\text{F}}{}_{\mu}.
\end{eqnarray}
As a rough guide, these operators have to be suppressed by about (4 TeV)$^2$ to be safe
\cite{Gregoire:2003kr,Barbieri:2000gf}.  To simplify the analysis, we will take the $SO(4)$ symmetric 
limit $g_L=g_R$.  In this restricted case, in order to get the correct $\sin \theta_W$
requires the relation $\sqrt{3}\theta' \approx \theta$ at small $\theta$'s.  
Thus, the $SU(2)$ operators
give the tightest bound.  Of these, the Higgs-Fermi $SU(2)$
operator turns out to be the most constrained giving a constraint 
\begin{eqnarray}
\quad m_{W'} \gtrsim 1.8 \TeV.
\end{eqnarray}
For the value $f=700 \GeV$, this corresponds to a limit $\theta \lesssim 1/4$. 
However, to be safe we'll later take as a benchmark value $\theta' = 1/5\sqrt{3}, 
\theta \approx 1/5$ from which to compare with experiment.
Note that for this near-oblique limit to exist, it was crucial that 
the light generations could be decoupled from the heavy gauge bosons.  Again, only 
in this limit is an analysis of the oblique corrections S and T  meaningful.

\subsection{Custodial $SU(2)$}
\label{sec: tparameter}
Custodial $SU(2)$ violating effects are highly constrained by precision electroweak
tests and this model's primary motivation is to minimize any such violation.  
Custodial $SU(2)$ violation is conveniently parameterized by corrections to the 
$\rho$ parameter (or equivalently the T parameter).   In a little higgs
model, there are potentially
five sources of custodial $SU(2)$ violation.   
The first possible contribution is that of expanding out the kinetic term in terms of the
higgs field.  The non-linear sigma model kinetic term
contains interactions at high order that could give custodial $SU(2)$ violating masses
to the W and Z.  However, in this model there is no violation at any order.  
This is due to the fact that the kinetic term is invariant under a global 
$SO(4)_D$ that is broken down by the higgs vev to custodial $SU(2)$.  As a matter of fact,
all terms in the expansion of the kinetic term just shift the value of the higgs vev
$v$, which gives $\delta\rho = 0$. 

\subsubsection*{Vector Bosons}
\label{Sec: Vector Bosons}
The second possibility is that integrating out the TeV scale gauge bosons (the $W', B', 
W^{r\,\pm}$) can generate a custodial $SU(2)$ violating operator.  This is typically 
denoted as 
\begin{eqnarray}
\mathcal{O}_4 = |h^\dag D_\mu h|^2.
\end{eqnarray}
In all previous little higgs theories, integrating out the $W'$ gauge bosons does not generate this operator 
at $O(v^2/f^2)$ and this holds true for this model as well.  On the other hand, 
integrating out the $B'$ and $W^{r\,\pm}$ {\em does} generate
this operator, but with opposite sign!  There is a cancelation with the total contribution 
\begin{eqnarray}
\delta\rho_\text{Gauge Boson} = -\frac{v^2}{16f^2}\sin^2 2\theta'.
\end{eqnarray}
Note that as advertised this vanishes in the limit $\theta' \to 0$, which is the same limit where the standard
model fermions decouple from the $B'$.  At the benchmark values $\theta' = 1/5\sqrt{3}, f=700\GeV$,
this gives a contribution $\text{T}_\text{gauge} = -.056$.  One can see that 
the addition of the extra $W^{r\,\pm}$ gauge bosons has 
provided an extra suppression factor of $\sin^2 2\theta' \approx 1/20$.

\subsubsection*{Triplet Vev}
The third contribution to custodial $SU(2)$ violation comes from the triplet vevs.
The key point is that the potential (Eq. \ref{eq: Potential}) is custodial $SU(2)$ invariant
except for the $\Delta \lambda_\mathbf{3}$ term generated by the gauged $U(1)$.  
The non-oblique corrections already prefer small $\theta'$ and thus small $g_1$.  
Therefore custodial $SU(2)$ violation in the potential should be small, and we
should expect that $\Delta \lambda_\mathbf{3} \ll \lambda_\mathbf{3}^\pm$.
Calculating the triplet vev contribution, we find
\begin{eqnarray}
\delta\rho_\text{triplet} = \frac{v^2}{16f^2}\left[\left(\frac{\lambda_\mathbf{3}^- -\lambda_\mathbf{3}^+ -\Delta \lambda_\mathbf{3}}
{\lambda_\mathbf{3}^- +\lambda_\mathbf{3}^+ +\Delta \lambda_\mathbf{3}}\right)^2
- \left(\frac{\lambda_\mathbf{3}^- -\lambda_\mathbf{3}^+}{\lambda_\mathbf{3}^- +\lambda_\mathbf{3}^+}\right)^2 \right]
\approx \frac{v^2}{4f^2}\frac{\lambda_\mathbf{3}^- (\lambda_\mathbf{3}^+ -\lambda_\mathbf{3}^-)}
{(\lambda_\mathbf{3}^- +\lambda_\mathbf{3}^+)^3}\Delta \lambda_\mathbf{3}. 
\end{eqnarray}
where we have expanded to first order in $\Delta \lambda_\mathbf{3}/\lambda_\mathbf{3}^\pm$ to get the end result.
In comparison with the ``littlest higgs'', there is now a beneficial 
$\Delta \lambda_\mathbf{3}/\lambda_\mathbf{3}^\pm$ suppression.    
We cannot really say anything more in the effective field theory
since there are unknown order one factors in the relation between the $\lambda_\mathbf{3}$'s and the coefficients
as calculated in the Coleman-Weinberg potential.  However, to get a feel for the 
expected size of the contribution, we can take the Coleman-Weinberg coefficients at 
face value which for $y_1 = 2$, $g_L = g_R$, and $f = 700 \GeV$, gives
the plot T vs. $\theta'$ as shown in figure \ref{fig: tripletT}. 
\FIGURE{
\epsfig{file=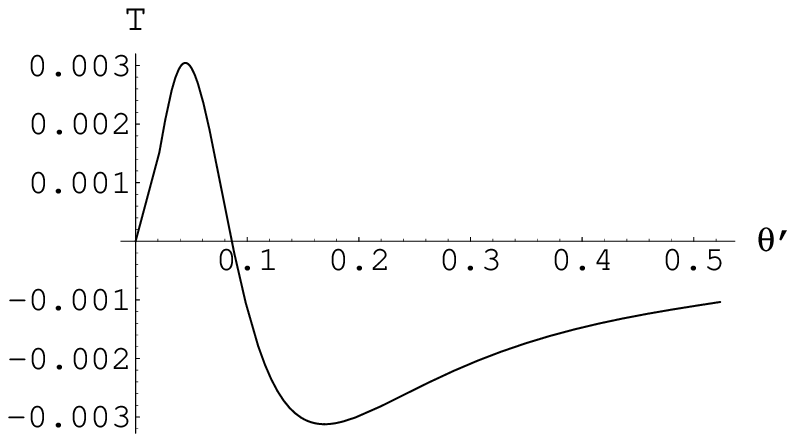}
\caption{Triplet Contribution to T as a function of $\theta'$ with 
$y_1 = 2$, $g_L = g_R$, and $f = 700 \GeV$.}
\label{fig: tripletT}}
In the limit $g_L=g_R$, there is an upper bound $\theta' \lesssim \pi/5$ (in order
to get the correct $\sin \theta_W$) which is 
why the graph is cut off on the right.  Order one factors aside, 
it is obvious that the triplet contribution
to T is negligibly small due to the extra suppression described above.     

\subsubsection*{Top Sector}
The fourth source of custodial $SU(2)$ violation is the 
introduction of new fermions in the top sector.  To calculate the effects of the
extra fermions, it is easiest to compute the contributions to 
T through vacuum polarization diagrams by the definition
\begin{eqnarray}
\text{T} \equiv \frac{e^2}{\alpha \sin^2\theta_W \cos^2\theta_W m_Z^2}[\Pi_{11}(0)-\Pi_{33}(0)].
\end{eqnarray}
If we ignore the small mixing effects 
induced by the b quark mass, the contribution is parameterized by the 
single parameter $\theta_t$ where
\begin{eqnarray}
y_1 = \frac{\sqrt{2}y_t}{\sin\theta_t}\hspace{0.5in}
y_2 = \frac{\sqrt{2}y_t}{\cos\theta_t}.
\end{eqnarray}
In figure \ref{fig: Tops}, plots of the T contribution versus $\theta_t$ are 
plotted for $f$ = 700 GeV and $f$ = 900 GeV, centered around $\theta_t = \pi/4$.  
Note that the standard model contribution to T has already been subtracted off from the total
top sector contribution, in order to give the final plotted results.
A good fit to the T contribution in the range of $\theta_t$ plotted is
$\cos^2{\theta_t} \cot{\theta_t}$, where the fit gets bad in the $\theta_t \leq 1/2$ region.  
As we change $f$, the constant of proportionality roughly scales as $1/f^2$.
In figure \ref{fig: mtprime}, the dependence
of $m_{t'}$  on $\theta_t$ is also plotted.  An important point is that
naturalness puts an upper bound
constraint on the mass $m_{t'}$.  By the standard given in \cite{Arkani-Hamed:2002qy}, 
for a 200 GeV Higgs, 10\% fine-tuning restricts $m_{t'} \lesssim $ 2 TeV.  
Fortunately, as the figures show, it appears possible to get corrections to T within the 1-$\sigma$
bound at $f$ scales consistent with this amount of fine tuning.  
It is also important to keep in mind that these T contributions are quite mild (this appears to 
be a generic feature of little higgs models).  For instance, 
the standard model
top quark contribution is $\text{T} \approx 1.2$ which is quite larger than the largest value in 
the 
plot of $0.35$.  
It is also well known that moderate positive values of T increase the upper bound on the 
higgs mass \cite{Peskin:2001rw}.   Since $\theta_t$ will be varied
during the fit, this will dramatically change the allowed higgs masses.   

\DOUBLEFIGURE{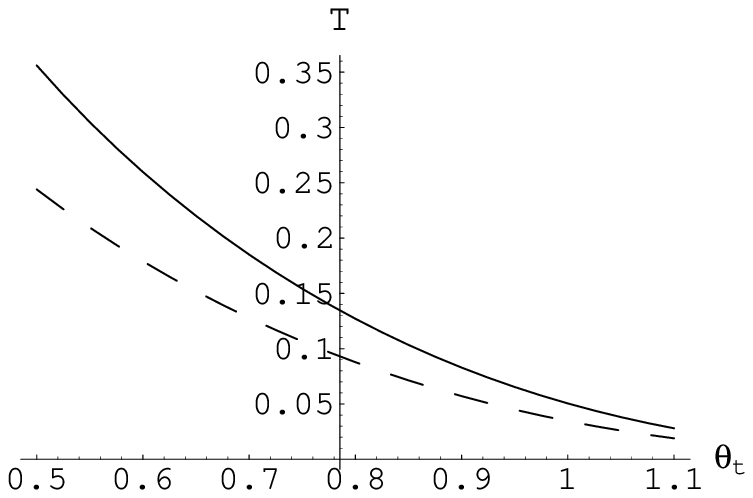}{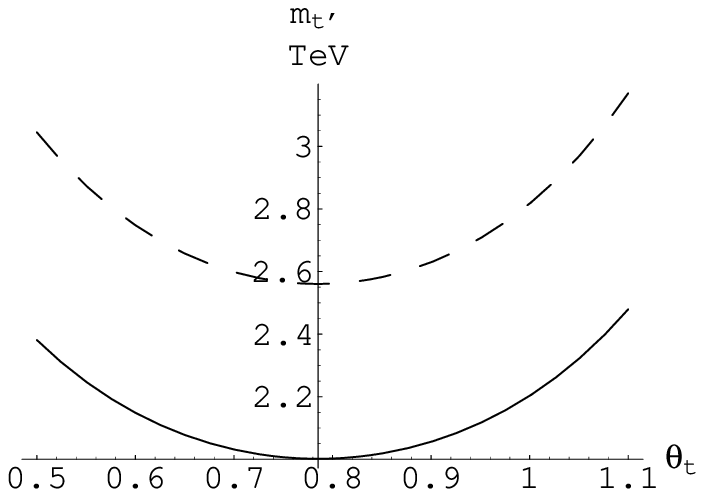}
{Top Sector Contribution to T as a function of $\theta_t$.
The solid line is for $f$ = 700 GeV while the dashed line is for
$f$ = 900 GeV.\label{fig: Tops}}
{The heavy top mass $m_{t'}$ 
as a function of $\theta_t$.  Again the solid line is for 
$f$ = 700 GeV while the dashed line is for $f$ = 900 GeV.
\label{fig: mtprime}}

\subsubsection*{Higgs}
Finally, the higgs itself will contribute to T.  The contribution is well known and 
we will use the explicit formula contained in \cite{Hagiwara:1994pw}.
For the purposes of this paper, we will take the S,T origin when $m_\text{higgs, ref}$ = 115 
GeV.  As we increase the higgs mass, this T contribution gets large and negative.

\subsection{S Parameter}
\label{sec: sparameter}
The S parameter along with the T parameter gives a good handle on the oblique 
contributions of
any new physics.  To the order at which we have been calculating
(i.e. $v^2/f^2$), there are only two sources of S contributions.  The first contribution is
that of the higgs.  
Again, we use the result in \cite{Hagiwara:1994pw}.  
This gives a positive S contribution for higgs masses larger than our chosen reference mass.

\FIGURE{
\epsfig{file=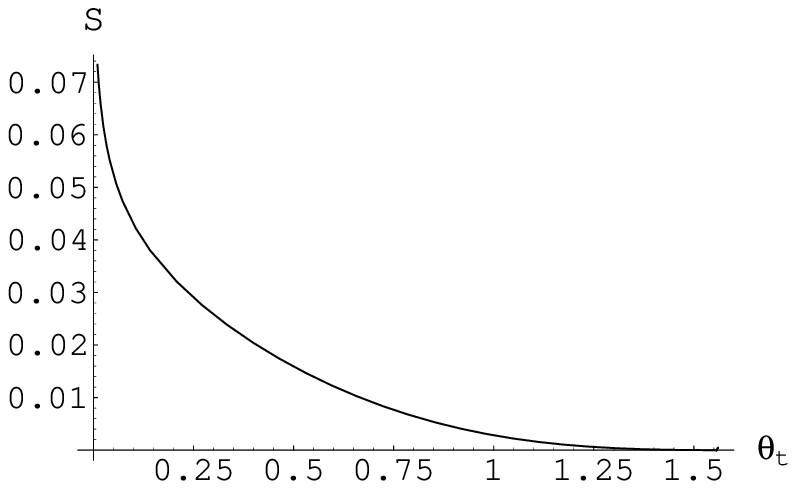}
\caption{Top Sector Contribution to S for $f$ = 700 GeV as a function of $\theta_t$.}
\label{fig: stop}}

The second contribution to S comes from the extended top sector and again is best calculated
via vacuum polarization diagrams using the definition
\begin{eqnarray}
\text{S} \equiv -16\pi \Pi'_{3Y}(0).
\end{eqnarray}
In figure \ref{fig: stop}, we have plotted the beyond the standard model 
S contribution from the top sector for $f$ = 700 GeV.  For the region of naturalness,
the contributions to S are quite small and do not measurably affect the fit of the model. 

\subsection{Summary of Limits}
\label{Sec: Summary}
\FIGURE{
\epsfig{file=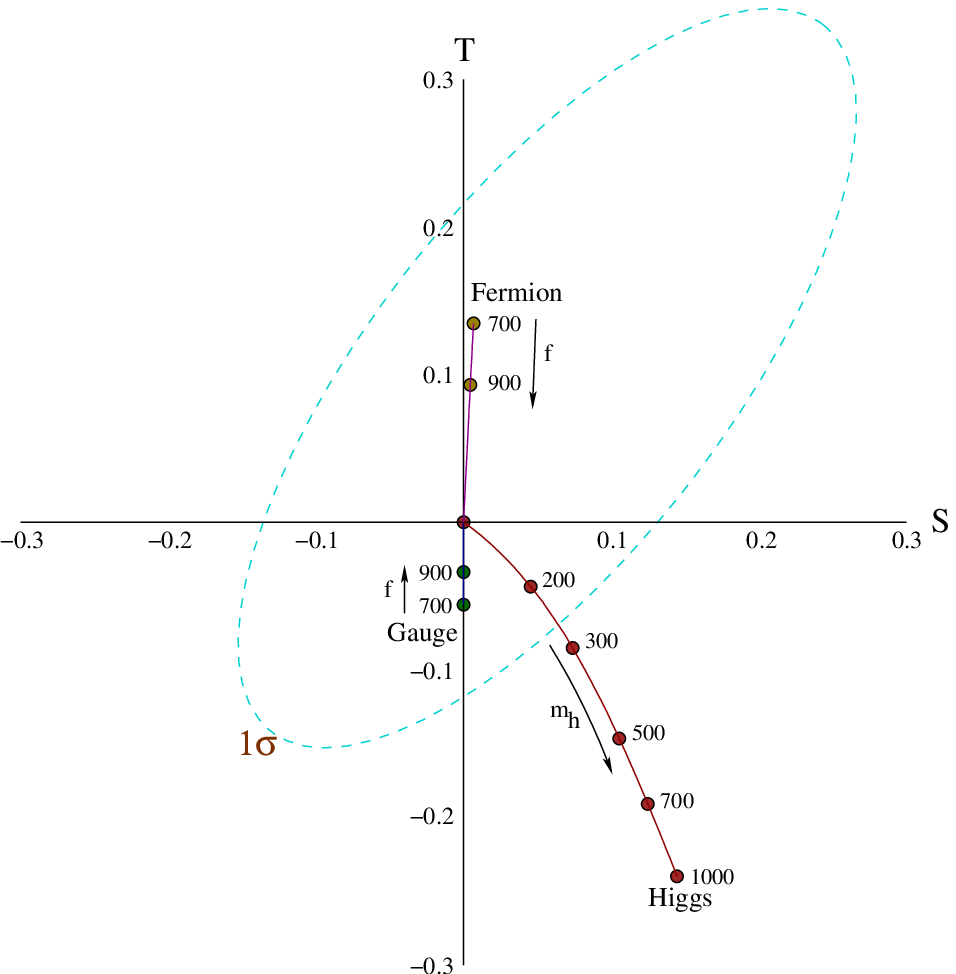}
\caption{The approximate 1$\sigma$ ellipse in the S-T plane.  The origin (S=0,T=0) corresponds
to the reference values $m_h =115 \GeV$ and $m_t = 174.3$ GeV.  The higgs contribution
for increasing higgs mass is plotted for the values (115,200,300,500,700,1000) GeV 
which slopes down and to the right.  Two representative points of the extra fermion and gauge
contributions have also been plotted for $\theta_t=\pi/4$, $\theta'=1/5\sqrt{3}$ and $f=700,900$ GeV.}
\label{fig: st}}

Now, the fit to S and T can be performed.  
In figure \ref{fig: st}, the approximate 1$\sigma$ ellipse in the S-T plane as given in 
\cite{Hagiwara:fs} has been plotted.
Note that the (S=0,T=0) origin has been set to the reference
values $m_h = 115$ GeV
and $m_t = 174.3$ GeV.  Sloping down and to the right, the exact contribution due to the higgs has been plotted for the
masses $m_h=(115, 200, 300, 500, 700, 1000)$ GeV.  
To represent the other two contributions, two points of the beyond the standard model
fermion and gauge contributions have also 
been plotted for the values $\theta_t = \pi/4$, $\theta' = 1/5\sqrt{3}$, 
and $f=700,900$ GeV. The fermionic contribution
generically points up and slightly to the right whereas the gauge contribution points
downward.  To find where the model is on the S-T plane,
these three contributions should be added.

To be specific, we'll focus on the value $f=700$ GeV as this limits the amount of fine-tuning in the
model.  In figure \ref{fig: stadded}, the S and T contributions for the higgs and
fermions are summed for $\theta_t=\pi/4$ and $\theta' =1/5\sqrt{3}$.  From the graph there appears
to be a generous range of $m_h$ that falls within the 1$\sigma$  
limits, at least $115 \GeV \leq m_h \lesssim 400 \GeV$.  If $\theta_t$ is changed, 
larger higgs mass can be attained.  Although changing $\theta_t$ from the 
equal mixing value $\pi/4$ increases fine-tuning, as $m_h$ increases the 
higgs mass parameter also increases which will reduce the fine-tuning, and thus
$\theta_t$ can be manipulated as the higgs gets heavier.  This freedom helps
since reducing $\theta_t$ will increase the fermion contribution to T (and only slightly increase S) 
as required to stay within the ellipse  \cite{Peskin:2001rw}.  For instance, at $\theta_t = 
\pi/6$, we can still tolerate a 1 TeV higgs mass.  
Changing $\theta'$ produces less of an effect,
but as it goes to zero, it can also help improve the fit at large higgs mass. 
Of course, at higgs masses about a TeV, the little higgs mechanism is not even required if the  
cutoff is taken to be 10 TeV.  
However, within our model, we see that a large region of parameter space is allowed by 
both 
the S-T fit and naturalness. In figure \ref{fig: mhvsthetat}, $\theta'$ and $f$ 
have been fixed while $\theta_t$ and $m_h$ are scanned; all points in the shaded region 
fit within the $1\sigma$ S-T ellipse, while all points above a dashed line are 
consistent with that percentage of fine-tuning 
(again using the fine-tuning definition of \cite{Arkani-Hamed:2002qy}).
There is quite a large range of higgs masses
allowed by precision constraints, and most of it is within ten percent fine-tuning or better.

\DOUBLEFIGURE{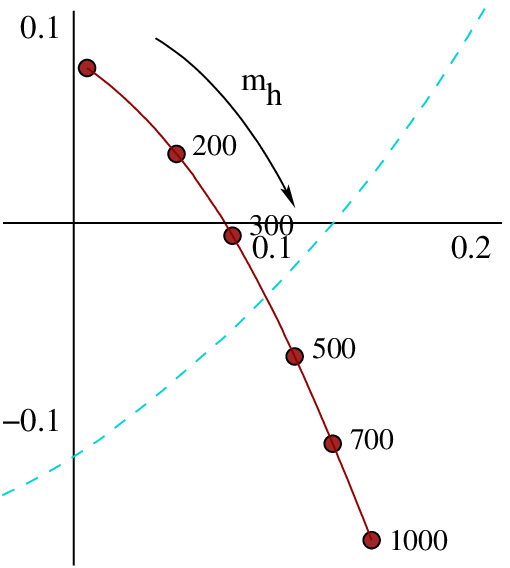}{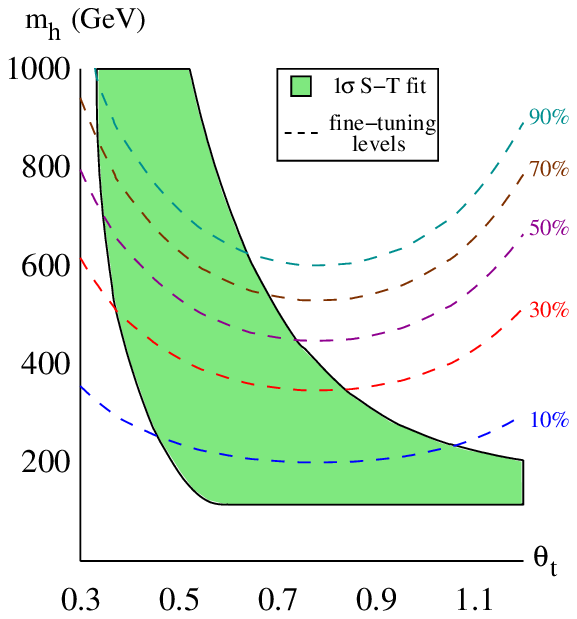}
{A closeup of the S-T plot with the summed contributions of the higgs, gauge bosons, 
and fermions.  In this plot, 
the values $f = 700 \GeV$, $\theta_t=\pi/4$, $\theta' = 1/5\sqrt{3}$ are fixed, but higgs mass
is allowed to vary. \label{fig: stadded}}
{A scan of $m_h$ vs. $\theta_t$ with the fixed values $\theta' =1/5\sqrt{3}$
and f= 700 GeV.  The shaded region is allowed by the $1\sigma$ fit while points above
a dashed line are consistent with that percentage of fine-tuning.
\label{fig: mhvsthetat}}

As a brief comment on more general $f$ values, 
the positive fermion contribution to T decreases as $f$ increases at a given $\theta_t$,
so it is more difficult to get within the ellipse for very large higgs mass at large $f$.  For instance,
going up to $f=900 \GeV$ pushes the range for $\theta_t=\pi/4$ and $\theta'=1/5\sqrt{3}$ 
down to about $115 \GeV \leq m_h \lesssim 350 \GeV$.
However, it is our hope that naturalness will help to keep $f$ low, so that the TeV scale particles can still 
be discovered at the LHC.  

Two more comments on this fit should be made.  First of all, the experimental error in
the top mass gives an uncertainty in the standard model contribution to S and T.  With the
current error of $\pm 5$ GeV, this introduces an unknown $\pm .07$ contribution to T (the 
change in S is small), which can significantly affect the fit.  
The other thing to note is to remember
that $O(v^4/f^4)$ effects and higher have been neglected.  For instance, as seen in 
\cite{Gregoire:2003kr},
S contributions from $O(v^4/f^4)$ and dimension 6 operators suppressed by $\Lambda^2$ 
are of the order $\pm .02$.  Thus, to go beyond the $O(v^2/f^2)$ analysis as presented
here will require some assumptions about the UV completion. 

As a conclusion to this section, we plot some sample spectrums for an allowed region of 
parameter space in figure \ref{fig: spectrum}, with $f=700$ GeV.  
For the heavy quark sector, we have 
allowed $\theta_t$ to vary.  The $t'$ quark is the heavy charge 2/3 quark 
that cuts off the top
yukawa quadratic divergence, and is nearly degenerate with the charge $-1/3$ $b'$ quark 
(the $b'$ is usually about 5-10 GeV heavier).  
The $T$ and $\Psi$ quark (charge 2/3 and 5/3 respectively) 
are exactly degenerate at tree level and are
not important in cutting off the top quadratic divergence.  In general, the $(T,\Psi)$ pair
is lighter than the $(t',b')$ pair.
In the heavy gauge boson sector, the $W'$ is generally the heaviest and the $B'$ and $W_r^\pm$
are nearly degenerate (with the $B'$ heavier).  As $\theta$ decreases, all of the states get heavier
and more degenerate.
Finally, in the scalar sector, the simplifying assumption that naturalness in the gauge contribution sets
$\lambda_\mathbf{1}^\pm = 3\lambda_\mathbf{3}^\pm$ has been assumed.  This naturalness condition sets the scalars to be heavier
than the triplets by a factor of $\sqrt{3}$.  To simplify it even further, we've also 
assumed that $\lambda_\mathbf{3}^- =\lambda_\mathbf{3}^+$ and picked a higgs mass of 200 GeV.
All these particles have TeV scale masses and should be searched for at the LHC.  

\FIGURE{
\epsfig{file=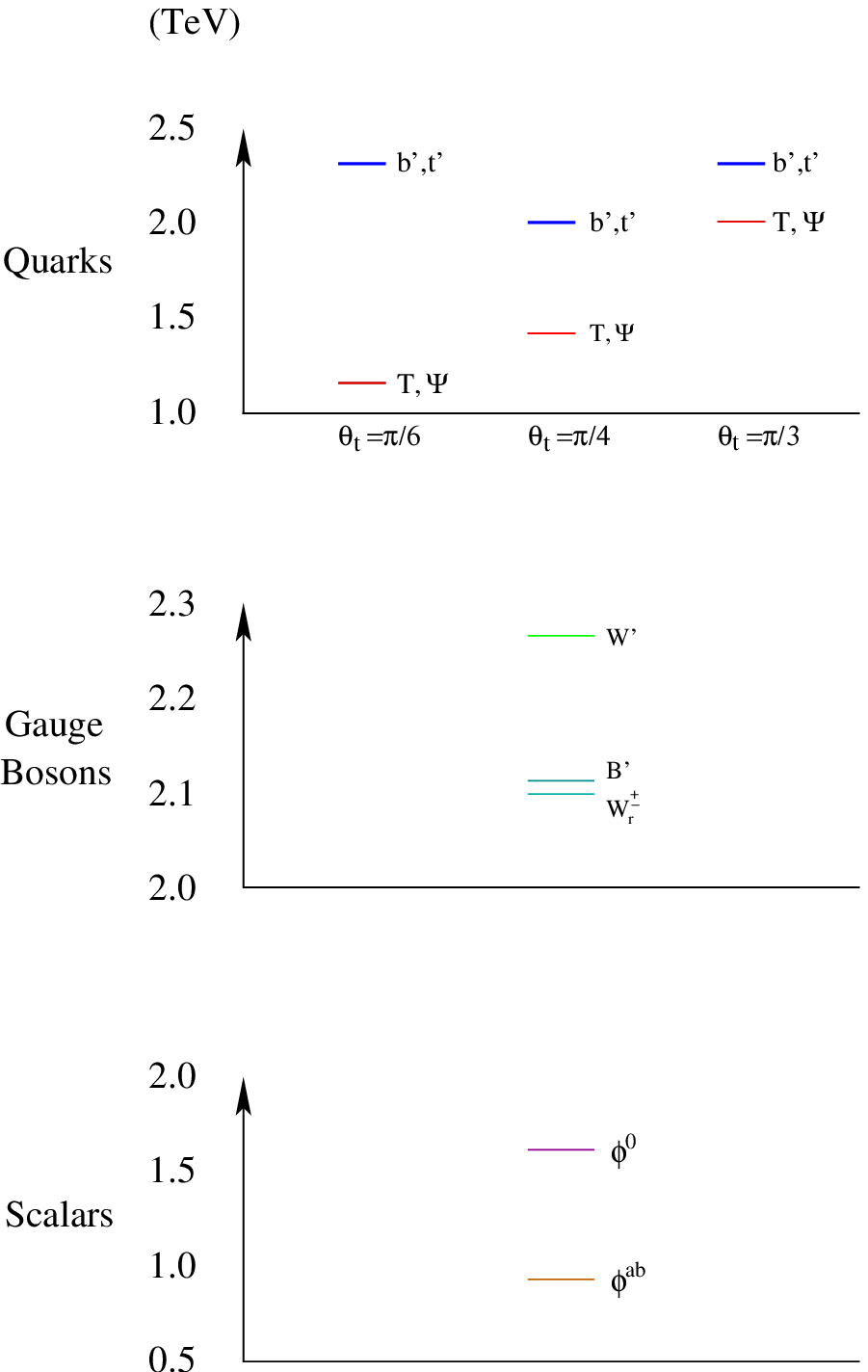}
\caption{A sample spectrum for the values $f= 700 \GeV, \theta =1/5,
\theta'=1/(5\sqrt{3})$.  In the scalar sector, $m_h = 200 \GeV$ and 
$\lambda_\mathbf{1}^\pm =3\lambda_\mathbf{3}^\pm = 4/3$ has been assumed.}
\label{fig: spectrum}}

\section{Conclusion}
\label{Sec: Conclusion}
In this paper, a new ``littlest higgs'' model with custodial $SU(2)$ symmetry has been
analyzed.  Precision electroweak analyses of little higgs models has given suggestions
on what features little higgs theories should realize, and with this motivation, the 
$\frac{SO(9)}{SO(5)\times SO(4)}$ model has been proposed in order to be easily compatible
with precision constraints.  Some of the unique features of the model include:
\begin{itemize}
\item Psuedo-goldstone bosons with custodial $SU(2)$ preserving vevs, comprised of a single light higgs doublet and at the TeV scale, a singlet and three $SU(2)_W$ triplets, similar
to \cite{Georgi:1984af}.
\item At $O(v^2/f^2)$, the $1\sigma$ S-T fit allows a generous range of higgs masses 
 in a large region of parameter space that is consistent with naturalness.  
\end{itemize}
The other features follow that of the original ``littlest higgs'', including the 
generation of the higgs potential through gauge and fermion interactions as well as 
the fermion sector driving electroweak symmetry breaking.  At low energies, the effective
theory is the standard model, with extra states at the TeV scale to cut off the quadratic
divergences to the higgs.  Once again, we emphasize that 
the precision constraints are mild and a large region of higgs masses is allowed in parameter
space where the higgs mass is natural.

In analyzing the one loop quadratically divergent term in the Coleman-Weinberg potential,
we discovered that the gauge interactions introduced a saddle point instability in the vacuum.
Two solutions to stabilize the vacuum were presented, either through extending the top 
sector or writing down operators that could give same sign contributions to the singlet
and/or triplet masses.  As an aside, we mention here briefly two ``littlest higgs'' models 
that also contain custodial $SU(2)$ where the preferred vacuum is stable.  Firstly, changing 
the global symmetry from $SO(9)$ to $SU(9)$ changes the breaking pattern to 
$SU(9) \to SO(9)$.  In this model, the upper $SU(4)$ gauge group can be gauged instead of
$SO(4)$.  The $SU(4)$ gauge interactions prefer the off-diagonal vacuum and stabilize the
$(\phi-X)^2$ terms.  This along with the top sector given earlier can stabilize the 
vacuum.  This theory contains 2 higgs doublets, 3 singlets, and 6 triplets and should 
preserve custodial $SU(2)$ in the same way as the $SO(9)$ model.  

Recently, the idea of UV completing the ``littlest higgs'' via strong interactions giving
rise to composite fermions and composite higgs was introduced \cite{Nelson:2003aj}.
This idea requires the top sector to be comprised of full multiplets of the global symmetry.
A model with custodial $SU(2)$ symmetry that conceivably 
could be UV completed in this manner is one based on the coset $\frac{SU(8)}{Sp(8)}$,  
where the upper 4 components of the 8 is $4 \equiv (2_L + 2_R)$ and $SU(2)_L\times SU(2)_R \times SU(2)\times U(1)$
is gauged.  This time the gauge interactions naively make the vacuum a local maximum, but 
the sign could depend on the UV completion and is just a discrete choice.  The spectrum of
this theory turns out to be 4 higgs doublets and 5 singlets! However, as one can see from 
these other examples, the model presented in this paper has the simplest spectrum, displays 
all the important physics, and is a complete and realistic model.

In summary, little higgs models are exciting new candidates for electroweak symmetry
breaking.  They contain naturally light higgs boson(s) that appear as pseudo-goldstone bosons
through the breaking of an approximate global symmetry.  With perturbative physics at the
TeV scale, these models produce relatively benign precision electroweak corrections.  In
this paper, one model that realizes custodial $SU(2)$ symmetry has been described, which 
may give some insight into why the standard model has worked so well for so long.  In the near
future, experiments at the LHC should start giving indications whether or not 
these candidate theories play a role in what comes beyond the standard model.      

\section*{Acknowledgements}
We would like to thank Nima Arkani-Hamed and Thomas Gregoire 
for many helpful discussions and especially Jay Wacker both for 
discussions and for suggesting the 
beginning framework of the model.  We also thank our anonymous
JHEP referee for many insightful comments.    
This work was supported by an NSF graduate student fellowship.

\appendix
\section{Generators and Notation}
\label{Sec: Gens}
The $SO(4)$ commutation relations are:
\begin{eqnarray}
[T^{mn},T^{op}] = \frac{i}{\sqrt{2}} ( 
\delta^{mo} T^{np}
-\delta^{mp} T^{no}
-\delta^{no} T^{mp}
+\delta^{np} T^{mo})
\end{eqnarray} 
where $m,n,o,p$ run from $1,\ldots, 4$.  These generators can be
broken up into
\begin{eqnarray}
\nonumber
T^{l\,a} = \frac{1}{2\sqrt{2}}\epsilon^{abc} T^{bc} + \frac{1}{\sqrt{2}}T^{a4}
\hspace{0.5in}
T^{r\,a} = \frac{1}{2\sqrt{2}} \epsilon^{abc} T^{bc} - \frac{1}{\sqrt{2}}T^{a4}\\
\end{eqnarray}
where $a, b, c$ run from  $1, \ldots, 3$.
The commutation relations in this basis of $SO(4)$ are equivalent to 
$SU(2)_L \times SU(2)_R$:
\begin{eqnarray}
\nonumber
[T^{l\,a}, T^{l\,b}] = i \epsilon^{abc} T^{l\,c}, 
\hspace{0.5in}
[T^{r\,a},\tau^{r\,b}] = i \epsilon^{abc} T^{r\,c}, 
\hspace{0.5in}
[T^{l\,a}, T^{r\,b}] = 0.
\end{eqnarray}

\subsubsection*{Vector Representation}

The vector representation of $SO(4)$ can be realized as:
\begin{eqnarray}
T^{mn}{\,}^{op} = \frac{-i}{\sqrt{2}}( \delta^{mo} \delta^{np} - \delta^{no} \delta^{mp})
\end{eqnarray}
where $m,n,o,p$ again run over $1, \ldots, 4$ and $m,n$ label the 
$SO(4)$ generator while $o,p$ are the indices of the vector representation.
In this representation:
\begin{eqnarray}
\Tr\; T^A T^B = \delta^{AB}.
\end{eqnarray}

\subsubsection*{Higgs Representations}

For the higgs doublet, we have three equivalent forms of the representation.  First, 
there is the $SO(4)$ vector representation, denoted as
\begin{eqnarray}
\vec{h} \equiv \left(\begin{array}{c} h^a \\ h^4 \end{array}
\right) \text{where } a=1,2,3,
\end{eqnarray}
the $SU(2)_W$ doublet representation (with Y = $-\half$)
\begin{eqnarray}
h \equiv \frac{1}{\sqrt{2}}\left(\begin{array}{r}
 		h^4 + i h^3 \\
		-h^2 + i h^1 \end{array}\right)
\end{eqnarray}
and the two by two matrix
\begin{eqnarray}
H \equiv   (h^4 + \sigma^a \, h^a)/\sqrt{2} = \left(\begin{array}{cc}
 		 h &
		 -\epsilon h^* \end{array}\right)
\end{eqnarray}
where the antisymmetric tensor $\epsilon = i \sigma^2$.  
Under $SU(2)_L \times SU(2)_R$, this matrix transforms as
\begin{eqnarray}
H \to L\, H\, R^\dag
\end{eqnarray}
and thus a vev in the $h^4$ direction breaks $SU(2)_L \times SU(2)_R$ to 
custodial $SU(2)$.

The Coleman-Weinberg potential depends on the fields $\mathcal{H}^0$ and $\mathcal{H}^{ab}$
as defined by 
$\frac{1}{2f} \vec{h}\: \vec{h}\,^T = \mathcal{H}^0 + 4\mathcal{H}^{ab}\: T^{l\,a} T^{r\,b}$.
These are given in terms of the higgs fields as:
\begin{eqnarray}
\mathcal{H}^0 = \frac{1}{4f}|h|^2 \hspace{0.5in}
\mathcal{H}^{ab} = \frac{1}{8f}\left[\left(h^c h^c-h^4 h^4\right)\delta^{ab}-2h^a h^b
-2\epsilon^{abc}h^c h^4 \right]
\end{eqnarray}

\subsubsection*{Singlet and Triplet Representations}
In this theory, there are TeV scale scalars transforming as a singlet and as
triplets under $SU(2)_W$, which appear in the symmetric product of two $SO(4)$ vectors, 
{\it i.e.} $(\mathbf{4} \times \mathbf{4})_S =  \mathbf{1} + \mathbf{9} = (\mathbf{1}_L,\mathbf{1}_R) +
(\mathbf{3}_L,\mathbf{3}_R) $.
In the non-linear sigma model field $\Sigma$, these appear in the symmetric 
four by four matrix $\Phi$ and can be written as
\begin{eqnarray}
\Phi = \phi^0 + 4\phi^{ab}\: T^{l\,a} T^{r\,b}.
\end{eqnarray}
Note that since the left and right generators commute, this is a symmetric matrix.
These fields are canonically normalized and for the triplets, 
$SU(2)_W$ acts on the $a$ index in the triplet representation and $U(1)_Y$ acts on 
the $b$ index by $T^{r\,3}$ in the triplet $SU(2)_R$ representation.

\subsubsection*{Fermion representation} 
The $SO(4)$ vector representation fits nicely for the scalars, but is a bit cumbersome 
for the fermion sector.  However, taking inspiration from the $\vec{h}, H$ transformation
properties, it isn't hard to see the correct correspondence.  
Let's first start with a set of doublets $q_1,q_2$ in a two by two matrix 
\begin{eqnarray}
Q \equiv \left(q_1 \; q_2\right)
\end{eqnarray}
which transforms under $SU(2)_L \times SU(2)_R$ as 
\begin{eqnarray}
Q \to L\, Q\, R^{\dag}. 
\end{eqnarray}
Thus the $q$'s are $SU(2)_L$ doublets and the $SU(2)_R$ rotates them into each other.
Now, Q can be transformed into an $SO(4)$ vector by tracing
\begin{eqnarray}
\vec{q}\,^T = (q^a,q^4) \equiv \frac{1}{\sqrt{2}} \left(\Tr(-i\sigma^a Q),\Tr(Q)\right)
\end{eqnarray}
where $\vec{q}$ transforms under the $T^l, T^r$ generators of the $SO(4)$ representation. 
The normalization out front is important if this is to be completed into an $SO(5)$ vector 
by the addition of a singlet fermion $\Psi$ (this is just in order to keep canonical normalization under
group action).  Finally, the generalization of this correspondence to a fermion transforming under 
under an $SU(2) \times U(1)$ gauge group is straightforward.

\bibliographystyle{JHEP}

\end{document}